# The Axis and the Perimeter of the Temple

Subhash Kak

Canonical texts describing the plan of the Hindu temple allude to its astronomical basis, and in this Indian sacred geometry is not different in conception from the sacred geometry of other ancient cultures,[1] although it has its own unique features. If astronomical alignments characterize ancient temples of megalithic Europe, Egyptians, Maya, Aztecs, and Javanese, they also characterize Indian temples. In some temples, the *garbhagṛha* (innermost chamber) is illuminated by the setting sun only on a specific day of the year, or the temple may deviate from the canonical east-west axis and be aligned with a *nakṣatra* (constellation) that has astrological significance for the patron or for the chosen deity of the temple.

A part of the astronomical knowledge coded in the temple lay-out and form is canonical or traditional, while the rest may relate to the times when the temple was erected. The astronomy of the temple provides clues relevant not only to the architecture but also the time when it was built.

In this article, we consider the broadest design related to the sacred space associated with the Hindu temple. There is continuity in Indian architecture that goes back to the Harappan period of the 3$^{rd}$ millennium B.C.E., as described in Michel Danino's important work on the plan for the Harappan city of Dholavira.[2] For this reason, we devote our attention to the earliest description of the temple in Indian literature, which goes back to the Vedic period. For a background to the earliest Indian art and architecture the reader might refer to an earlier paper by the author.[3]

Specifically, we look at the astronomical significance of the lengths of the axis and the perimeter. Since sacred architecture often served as model followed in the design of cities (this is described, for example, by Volwahsen in his book *Cosmic Architecture in India*[3]), this question is of much relevance. The astronomical basis shown in this paper appears not to have been made note of in relatively recent temple architecture texts from India.

### Ritual and Plan of the Temple

The Agnicayana altar, the centre of the great ritual of the Vedic times that forms a major portion of the narrative of the Yajurveda, is generally seen as the prototype of the Hindu temple and of the Indian tradition of architecture (Vāstu). The altar is first built of 1,000 bricks in five layers (that symbolically represent the five divisions of the year, the five physical elements, as well as five senses) to specific designs. The Agnicayana ritual is based upon the Vedic division of the universe into three parts of earth, atmosphere, and sky (Figure 1), that are assigned numbers 21, 78, and 261, respectively; these numbers add up to 360, which is symbolic representation of the year. These triples are seen in all reality, and they enlarge to five elements and five senses in a further emanation.

The householder had three altars of circular (earth), half-moon (atmosphere), and square (sky) at his home (Figure 2), which are like the head, the heart, and the body of the Cosmic Man (puruṣa). In the Agnicayana ritual, the atmosphere and the sky altars are built afresh in a great ceremony to the east. The numerical mapping is maintained by placement of 21 pebbles around the earth altar, sets of 13 pebbles around each of the 6 dhiṣṇya (atmosphere for 13×6=78) altars, and 261 pebbles around the great new sky altar called the Uttara-vedi.

The Uttara-vedi is equivalent to the actual temple structure. Vāstu is the remainder that belongs to Rudra, and Vāstupuruṣa, the temple platform, is where the gods reside, facing the central square, the Brahmasthāna. Given the recursive nature of Vedic cosmology, it turns out that the Uttara-vedi also symbolized the patron in whose name the ritual is being performed, as well as puruṣa and the cosmos.

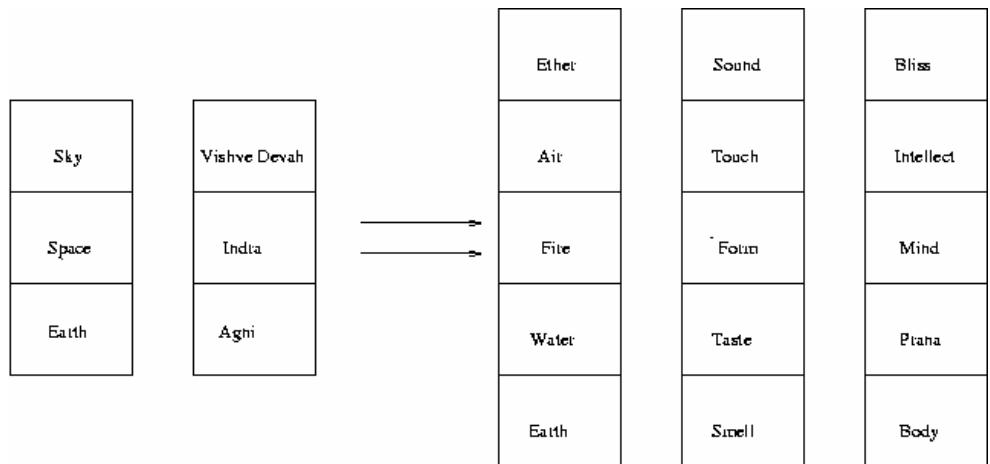

Figure 1: The tripartite division of the world and its enlargement into five layers

The underlying bases of the Vedic representation and ceremony are the notions of *bandhu-* (equivalence or binding between the outer and the inner), *yajña* (transformation), and *parokṣa* (paradox). To represent two more layers of reality beyond the purely objective, a sixth layer of bricks that includes the hollow svayamātṛṇṇā brick with an image of the golden puruṣa inside is made, some gold chips scattered and the fire placed, which constitutes the seventh layer (ŚB 10.1.3.7). The five layers are taken to be equivalent to the Soma, the Rājasūya, the Vājapeya, the Aśvamedha, and the Agnisava rites. The two layers beyond denote completion, since seven is a measure of the whole. The meaning of this is that the ceremonies of the great altar subsume all ritual.

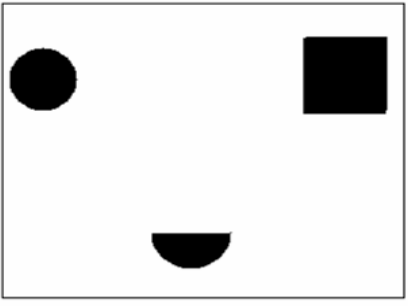

**Figure 2.** The three altars of the Vedic house: circular (earth, body), half-moon (atmosphere, prāṇa), square (sky, consciousness)

According to ŚB 7.4.1, within the hollow brick in the navel of the Uttara-vedi, a lotus-leaf is placed upon which is then placed a gold rukma (a disk, with 21 hangings), which symbolizes the sun. The golden puruṣa (representing Prajāpati as well as the Yajamāna himself) is laid on the back with the head towards the east on top the rukma.

On the sides of the golden puruṣa are two offering spoons, like two arms. Upon this image is placed the already mentioned svayamātṛṇṇā (self-perforated) brick. In total, there are three such bricks, in the centre of the first, the third, and the fifth layers. There are seven more bricks placed to the east of the svayamātṛṇṇā brick in the fifth layer. Next is a wooden mortar placed and on top of the mortar is placed the ukhā, the fire-pan which becomes the focus of the fire ceremony. The "sixth layer is the heavenly world, and the seventh layer is immortality" (ŚB 8.7.4.17-18).

That the worship of Śiva, Viṣṇu, and Śakti emerged from Agnicayana is described at length by Viśvambharanātha Tripāṭhī in his Agnicayana[5], which was published by Sampurnand Sanskrit University in Varanasi in 1990. The temple is not merely the buildings, the deity, but also the complex of the yajña, pūjā, or ceremonies performed there, so that in totality it represents both the *being* as well as the *becoming*. The becoming, or the transformation, requires the use of a special vocabulary related to inner processes. Briefly, Rudra is one of the names of Agni. According to Vājasaneyi S. 16.2, Agni has two forms, the auspicious Śiva and the fierce Rudra. During the building of the altar, Agni appears in its raudra manner, and to propitiate it the Śatarudriya homa is performed. This propitiation of Agni-Rudra is also done literally by a stream of water that drops out of an earthen pot hung over the liṅga. In one of the constructions of Nāciketa Agni, 21 golden bricks are placed one top of another to form the liṅga (Taittirīya Br. 3.1.1.6).

In the Vaiṣṇava tradition, the visualized golden puruṣa is Viṣṇu-Nārāyaṇa who emerges from the navel of the lotus on the Uttara-vedi that represents the waters, and for this reason is also called Padmanābha. The golden disk upon the lotus is then the sudarśana cakra of Viṣṇu.

In the building of the Uttara-vedi, seven special kṛttikā bricks (named ambā, dulā, nitatni, abhrayantī, medhayantī, varṣayantī, cupuṇīkā) together with aṣāḍhā, which is the eighth, are employed. The firepan (ukhā), which symbolizes Śakti, the womb of all creation, is taken to constitute the ninth. Nine represents completion (as well renewal, as in the very meaning of the Sanskrit word for nine, *nava*, or "new") and symbolizes the power of the Goddess. In later representation, which continues this early conception, the nine triangles of the Śri-Cakra represent Prakṛti (as a three-fold recursive expansion of the triples of earth, atmosphere, and the sky).

The ritual is sacred theatre[6] that describes dualities to help one transcend them. Rather than the indirect reference to two golden birds of the Ṛgveda, ŚB 10.5.2.9-11 says directly that there are two individuals within the body: the one in the right eye is Indra (representing articulated force) and the one in the left eye is Indrāṇī (Indra's consort representative of Prakṛti). "These two persons in the eyes descend to the cavity of the heart and enter into union with each other, and when they reach the end of their union, the individual sleeps."

But our objective here is not on the connections between Agnicayana and the traditional Śaiva, Vaiṣṇava, or Śākta systems. Rather, we wish to focus on the narrow question of the lay-out of the Agnicayana structures and relate it to the specifics of the dimensions of the axis and the perimeter of the classical Hindu temple. I have described the astronomy underlying Vedic ritual earlier[7] in several books and papers and it will not be further mentioned here.

### The Altar as a Representation of the Cosmos

The Śatapatha Br. informs us that the altar is to represent the mystery of time. Its dimensions are to represent earth. "As large as the altar is, so large is the earth" (ŚB 3.7.2.1) indicates that it symbolically represents objective knowledge. In the Śatapatha Śāṇḍilya says:

- Prajāpati is the year, and the bricks are the joints, the days and nights. The altar is the earth, the Agnicayana the air, and the *mahad uktham* the sky. The altar is the mind, the Agnicayana the air, and the *mahad uktham* the speech (ŚB 10.1.2.2-3).
- The Year, doubtless, is the same as Death. Prajapati said: "You do not lay down all my forms, making me either too small or too large. That is why you are not immortal... Lay down 360 enclosing stones, 360+36 yajuṣmatī (special) bricks, and 10,800 lokampṛnā (ordinary) bricks and you will be laying down all my forms, and you will become immortal." (ŚB 10.4.3.8)

> The 10,800 count represents the number of muhūrtas (48-minute interval) in a year.

The special yajuṣmatī bricks are placed 98 in the first layer, 41 in the second, 71 in the third, 47 in the fourth, and 138 in the fifth layer. These add up to 395; the earth filling between the bricks is taken to be the 396$^{th}$ brick. The sum of the bricks in the fourth and fifth layers together with one space filling is 186 (half the tithes in the solar year), the number of bricks in the third and the fourth layers equals one third the number of days in the lunar year, and so on.

Clearly, the objective is to represent the fact of the 360 divisions of the year (the additional 36 days represent the intercalary month) as well as other astronomical facts. The *bandhu-* relationship of the outer with the inner cosmos of the individual required an accurate representation of the outer so that a correspondingly accurate measure of the inner would become possible.

To understand how 10,800 lokampṛṇā bricks can fit the Uttara-vedi altar when the total number of bricks in the construction of the five-layered altar of 7½ square puruṣa altar is only 1,000, note that the ritual concludes at the end of the 95-year progressive enlargements of the altar by one square puruṣa per year. When we reach the altar of area 101½ square puruṣa at the end, the number of bricks it will require is:

$$1{,}000 \times 2/15 \times 203/2 = 13{,}533$$

Since not all the bricks are of the same size, it leaves room to place more than 3,000 smaller, suitably marked bricks on the fifth layer (ŚB 10.2.1.9-11). When the size of the altar was smaller, chanted meters could have substituted for the missing bricks.

Some of the ritual directly presents astronomical information as in the arrangement shown below which is described in the Śatapatha as representing the motion of the sun around the earth (the nākasads, ŚB 8.6.1). It is striking that this arrangement sees (accurately) the two halves of the year as being unequal by the use of 29 special bricks in the fifth layer of the altar.

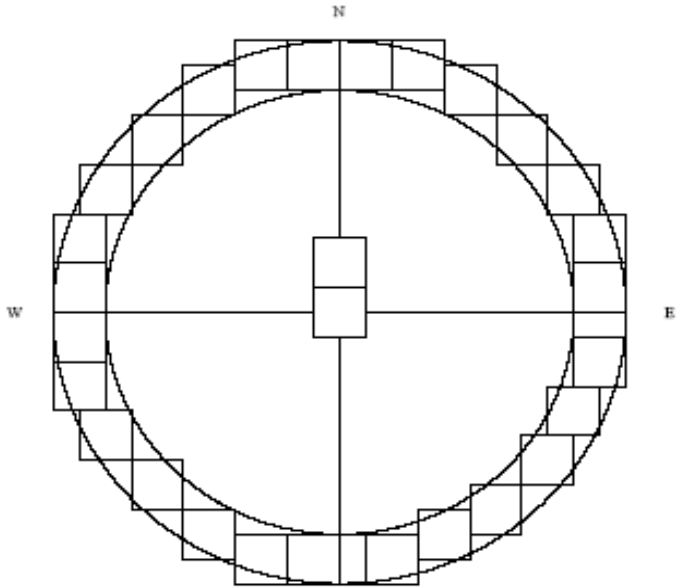

**Figure 3:** The representation of the sky in the fifth layer

In the Agnicayana ritual, the ritual was performed in a special area where first the three fires of the yajamāna are established in the west in an area called Prācīnavaṃśa, "Old Hall", or Patnīśālā, "Wife's Hall", whose dimensions are in the canonical ratio of 1:2.

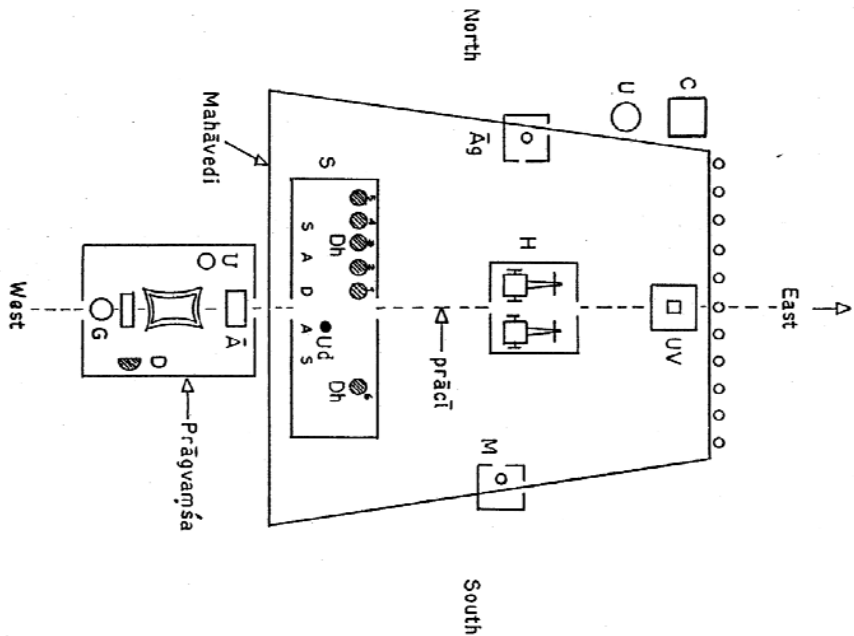

**Figure 4:** The ritual ground

The Prācīnavaṃśa (also called Prāgvaṃśa) has dimensions of 20×10 (Figure 4). Three steps (three puruṣa) from it to the east (ŚB 3.5.1.1) is the Mahāvedi, which is an isosceles trapezoid of spine 36 and the two sides of 30 and 24 units. The perimeter of the Mahāvedi is 126.25, whereas that of the Prācīnavaṃśa, taken separately is 60. But the Prācīnavaṃśa and the Mahāvedi are two components of the larger sacred ground and, therefore, they should be taken together. This unitary representation, in my view, is the plan of the prototype temple.

To see the significance of the plan, we now draw the Agnikṣetra within a rectangular area. It is appropriate here to be guided by the proportions that are clearly spelt out, such as that of 1:2 for the Prācīnavaṃśa, as also by numbers that are in terms of the metre numbers, which are used in a parallel representation of the altar. Amongst the metres, gāyatrī (24) is the head, uṣṇih (28) the neck, anuṣṭubh (32) the thighs, bṛhati (36) the ribs, paṅkti (40) the wings, triṣṭubh (44) the chest, and jagatī the hips; virāj (30) is invoked in the description of the Mahāvedi.

I think for accord with the measures which are multiples of 6, the left area was increased by an additional 1 puruṣas to the west to become 24×30 as in Figure 5, which is described as an appropriate proportion for a house in later texts such as Varāhamihira's Bṛhat Saṃhitā (53.4) indicating that it is an old tradition[8]. The Prācīnavaṃśa's share to the perimeter is 24+30+24=78, which is the atmosphere number. This is also in accord with the notion that the Prācīnavaṃśa is tripled in size in the completion of the Mahāvedi, going from 10×20 to 30×60.

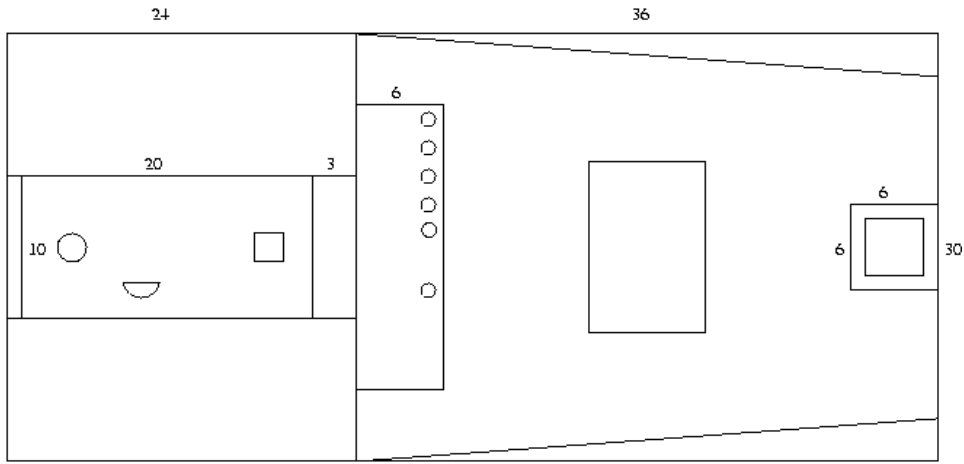

**Figure 5:** The temple plan

This is the basic temple plan, and it has the overall dimensions 60×30, with a perimeter of 180. The overall temple proportion of 1:2 is attested in later texts such as the Bṛhat Saṃhitā and Śilpa Prakāśa.[9]

In the Mahāvedi, first an area to make the six dhiṣṇya hearths under the Sadas, the shed or the tent, is marked at six puruṣa from the left (ŚB 3.6.1.3). To further east is Havirdhāna, the cart shed, and still further east, the Uttara-vedi, the great altar, in a square shape. One is enjoined to make the altar with each side the size of the yoke (ŚB 3.5.1.34), which is 86 aṅgulas (120 aṅgulas = 1 puruṣa), or in a measure of ten feet. Eggeling explains that there is disagreement regarding the location and size since there is another option "between four other measurements, viz. he may make it either one third of the area of the large altar, or of unlimited size, or of the size of the yoke or of tem of the sacrificer's feet" (Eggeling, vol II, page 119)[10].

I believe this ambiguity is deliberate since the location of the Uttara vedi would depend on its size, which is going to vary from 7½ square puruṣa to 101½ square puruṣa. In its basic placement, one would expect the determining factor to be the symmetry with the Sadas of 6 puruṣa width. This means that the Uttara-vedi will be built 54 puruṣas from the west, or 6 puruṣas from the east.

Thus the great square altar at the extreme east end of the Mahāvedi is marked off at a point which is 54 units away from the west end.

As the Agnicayana altars are made progressive larger by one square puruṣa each year in a 95-year sequence, symmetry requirements imply that the centre of the Uttara-vedi will come closer to the west. Therefore, in advanced constructions, the measure of 54 puruṣa separating the centre of the Uttara-vedi from the western edge will not hold. A few examples of the shapes of the Uttara-vedi are given in Figure 6.

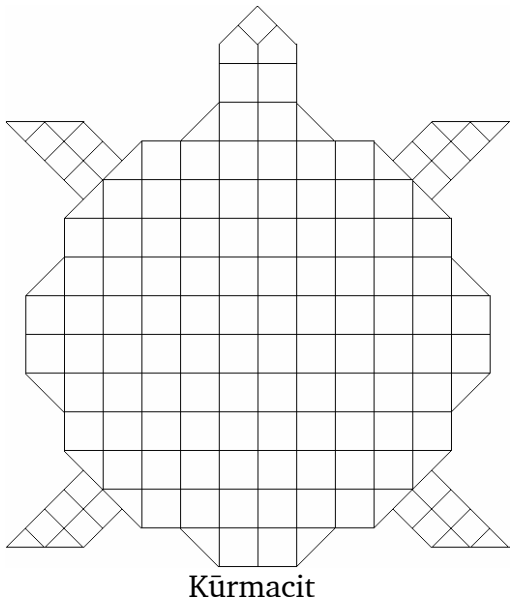

Kūrmacit

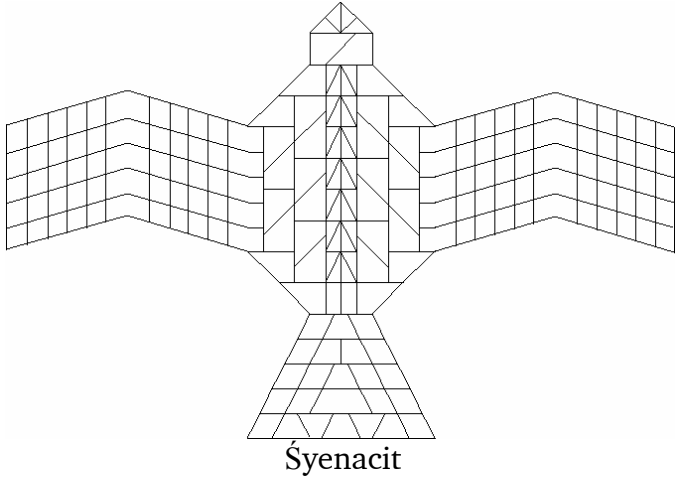
Śyenacit

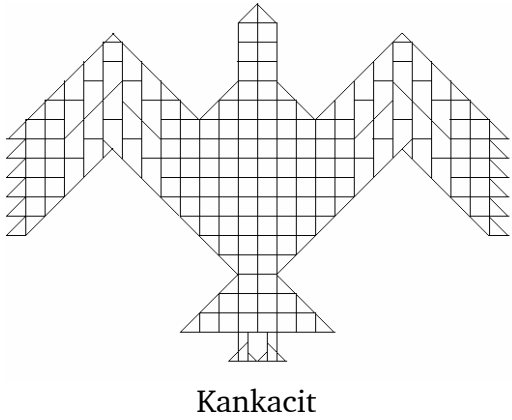
Kankacit

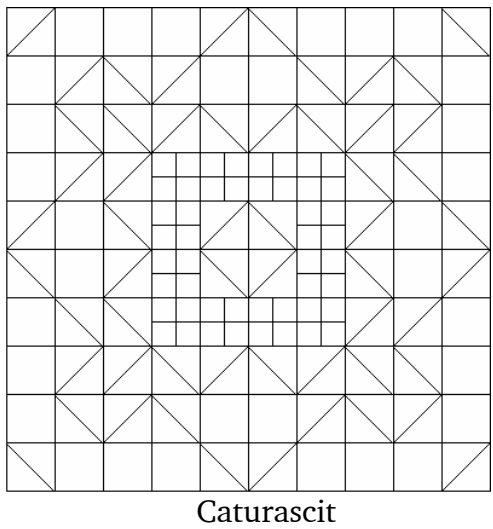
Caturascit

**Figure 6:** Four examples of Uttara-vedi: Kūrmacit, Śyenacit, Kankacit, Caturscit

Three of these represent time as turtle, eagle, and kite; in the later Caturascit, the representation is square, in the shape of the Vāstupuruṣa maṇḍala, and it is easy to see how that could be the prototype for the traditional temple plan. The Vāstupuruṣa is usually of 64 (or 81) squares in which the outer squares symbolize the 28 (or augmented 32) nakṣatras. The eight directions of space are presided over by 8 planets and 8 divinities of the nakṣatras; these squares, therefore, preside over the daily and the annual motions of the sun and the moon. Within the Vāstupuruṣa maṇḍala, twelve more assignments are made in the case of the 81-square plan for a total of 45 divinities. Utpala's commentary on the Bṛhat Saṃhitā 53.75 speaks of how the building should not face the corners of the square (of the cardinal directions) and how the direction chosen is related to the remainder when the perimeter is divided by 8, indicating the importance given to the perimeter.

### The Axis and the Perimeter

The Agnikṣetra or the later temple plan of the Vedic ritual represents two significant numbers, 180 and 54, which, when doubled, correspond to astronomical knowledge related to the 360 days of the year (attested in the Ṛgveda) and the ubiquitous number 108, which shows up as the number of beads in the rosary (japamālā), the number of dance movements (karaṇas) of the Nāṭya Śāstra, the names of the God and the Goddess, the number of pīṭhas, the number of dhāms, the number of arhats, and so on.

This number 108 has traditionally been derived from the auspicious number 9 when written as 1 + 8, from where it also become the auspicious number 18 of the number of Purāṇas, or the chapters of the Bhagavad Gītā. The number 108 is further seen as being auspicious since it is 27×4 where 27 is the number of nakṣatras. Some have pointed to its noteworthy number theoretic properties such as the symmetry in its representation as the product of the square of 2 with the cube of 3. The number 1,008, also considered auspicious, is viewed as the enlargement of 108 with the interposition of a 0, but this argumentation is incorrect since 10,008 is not an auspicious number. In truth, the auspiciousness of 1,008 is related to the fact that the Kalpa has 1,008 mahāyugas, and thus this number symbolizes completion of time.

In my *Astronomical Code of the Ṛgveda*, I argued that the correct interpretation of 108 is the distance to the sun and the moon from the earth in sun and moon diameters. This number codes a fundamental measure related to our physical universe. Figure 7 presents these proportions.

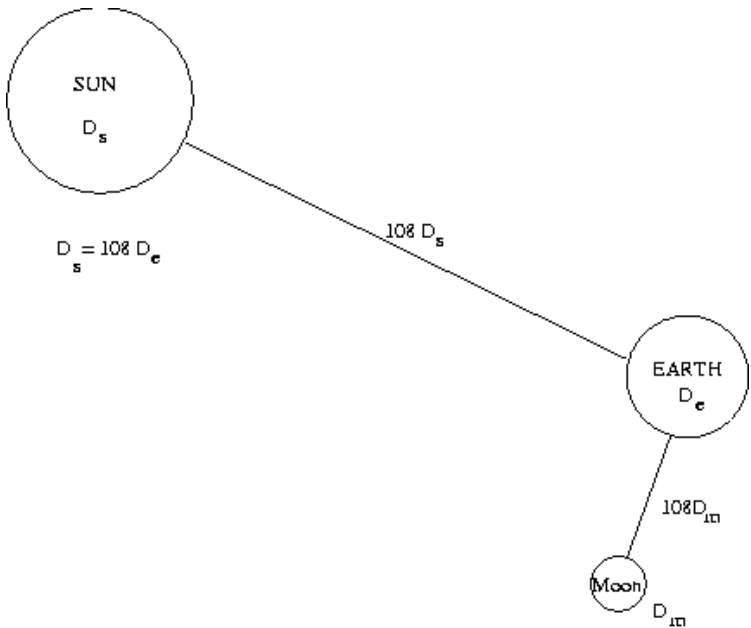

**Figure 7:** The earth, sun, moon system

The currently estimated mean diameter of the sun, $D_s$, is 1,392,000 km; the mean diameter of the earth, $D_e$, is 12,742 km; and the diameter of the moon, $D_m$, is 3,476 km. The estimated average distance between the earth and the sun is 149,600,000 km, so that

Average Distance to Sun/ Sun diameter ≈ 107.5

The distance between the earth and the moon varies considerably from the average perigee of 363,300 km to the average apogee of 405,500 km, for a mean of 384,400 km.

Average Distance to Moon/ Moon diameter ≈ 110.58

Diameter of Sun/ Earth diameter ≈ 109.24

It is thus correct for these distances to be approximated by the figure of 108. The origin of this number in ancient India may be the discovery that a pole of a certain height removed to a distance of 108 times its height has the same angular size as the sun or the moon. Therefore, the knowledge of the astronomical significance of this number in ancient India is not to be taken as anomalous. But since such a comparison made by the naked eye can only be correct to 1 or 2 percent, it seemed logical to take the number to be the round number 108.

This interpretation of 108 cannot be taken to be a coincidence since we also have the numbers 261 and 78 explicitly associated with the atmosphere and the sky.

Taken together, they add up to 339, which is approximately equal to 108×π, in accord with the notion that the sun, 108 units away from the earth, will inscribe 339 disks from rising to setting. These numbers also show up in the very organization of the Ṛgveda and other texts, as I have explained elsewhere, confirming their centrality in Vedic cosmological thinking.

This measure is also at the basis of the estimated distance to the sun in ancient canonical treatises. On the other hand, there is no evidence for the correct estimate of the size of the sun.

Discussion

The plan of the Hindu temple, as seen in its earliest form in the Agnicayana rite, is a representation of the cosmos. Its axis, from the western gateway to the *garbhagṛha* in the east, represents the distance to the sun (or the moon), and its perimeter represents the duration of the year in terms of the number 360. Specifically, the ratio 360/108 would characterize the standard temple in the proportions related to perimeter and axis.

Some later temples deviate from this standard in a variety of ways; in some, the perimeter is not the lunar year of 360 tithis (or civil year of as many days) but rather its count of 354 days. In others, the representation of the axis and the perimeter is not in linear measure but rather expressed symbolically.[11] Such deviations from the prototype make the layout unique and interesting and can tell us much about the cosmological ideas of its times.

The temple itself, in its three-dimensional form, codes several rhythms of the cosmos and specific alignments related to the geography of the place and the presumed linkages of the deity and the patron. The architecture may also incorporate themes related to royal power if it was built at the behest of a king.

But this does not mean that the relationship of the Hindu temple is to the physical cosmos alone. The Vedic philosophy of *bandhu* takes the numbers 360 and 108 to be central to the inner cosmos of the individual also. Therefore, walking 108 steps to the sanctum, or doing the 108 beads of the rosary, is a symbolic journey from the body to the heart of consciousness, which is the inner sun.

The Hindu temple has continued to reflect astronomical numbers and orientations, which is seen most dramatically in the great temple of Angkor Wat in Cambodia.[12]